\documentclass{iaus}
\usepackage{graphicx}

\newcommand{\araa}{\textit{ARA\&A}}
\newcommand{\apj}{\textit{ApJ}}
\newcommand{\apjl}{\textit{ApJ}}
\newcommand{\apjs}{\textit{ApJS}}
\newcommand{\mnras}{\textit{MNRAS}}
\newcommand{\aap}{\textit{A\&A}}
\newcommand{\nat}{\textit{Nature}}

\title[Chemical history of molecules in circumstellar disks] 
{Chemical history of molecules in circumstellar disks}

\author[R. Visser et al.]   
{Ruud Visser$^{1,2}$, Ewine F. van Dishoeck$^{2,3}$ \and Steven D. Doty$^4$}

\affiliation{$^1$Department of Astronomy, University of Michigan, 500 Church Street, Ann Arbor, MI 48109-1042, USA\\ email: {\tt visserr@umich.edu}\\[\affilskip]
$^2$Leiden Observatory, Leiden University, P.O.\ Box 9513, 2300 RA Leiden, the Netherlands\\[\affilskip]
$^3$Max-Planck-Institut f\"ur Extraterrestrische Physik, Giessenbachstrasse 1, 85748 Garching, Germany\\[\affilskip]
$^4$Department of Physics and Astronomy, Denison University, Granville, OH 43023, USA
}

\pubyear{2011}
\volume{280}  
\pagerange{xx--yy}
\jname{The Molecular Universe}
\editors{J. Cernicharo \& R. Bachiller, eds.}
\begin{document}

\maketitle

\begin{abstract}
The chemical composition of a protoplanetary disk is determined not only by in situ chemical processes during the disk phase, but also by the history of the gas and dust before it accreted from the natal envelope. In order to understand the disk's chemical composition at the time of planet formation, especially in the midplane, one has to go back in time and retrace the chemistry to the molecular cloud that collapsed to form the disk and the central star. Here we present a new astrochemical model that aims to do just that. The model follows the core collapse and disk formation in two dimensions, which turns out to be a critical upgrade over older collapse models. We predict chemical stratification in the disk due to different physical conditions encountered along different streamlines. We argue that the disk-envelope accretion shock does not play a significant role for the material in the disk at the end of the collapse phase. Finally, our model suggests that complex organic species are formed on the grain surfaces at temperatures of 20 to 40 K, rather than in the gas phase in the $T>100$ K hot corino.
\keywords{astrochemistry -- stars: formation -- circumstellar matter -- planetary systems: protoplanetary disks -- molecular processes}
\end{abstract}

\firstsection 
\section{Introduction}
We live on a world rich in complex organic molecules, all of which ultimately derive from a cold pre-stellar core with a much simpler chemical composition. How did the chemistry in the solar system evolve from that original simple state to its current complex state? This chapter focuses on one particular part of the entire evolutionary track: the transition from a pre-stellar core to a protoplanetary disk.

\begin{figure}[t]
\begin{center}
\resizebox{0.8\hsize}{!}{\includegraphics{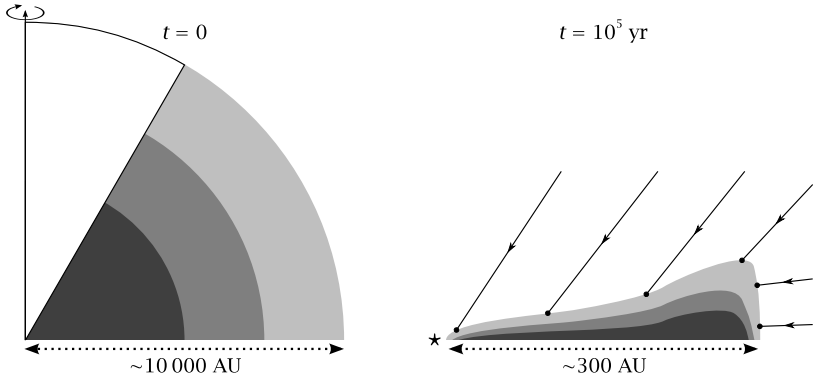}}
\caption{Schematic view of layered accretion. The outer parts of the original cloud core (left) end up as the surface layers of the circumstellar disk (right). The white part of the core is roughly the part that disappears into the outflow. The arrows illustrate the flow of material onto the disk. If the vertical extent of the disk is not taken into account properly, as happened in some older models, such streamlines would extend further in and the radii at which material accretes would be underestimated.}
\label{fig:acclayers}
\end{center}
\end{figure}

The chemical evolution in a collapsing envelope has been studied extensively with one-dimensional (1D) spherical models (e.g., \cite[Bergin \& Langer 1997]{bergin97b}; \cite[Rodgers \& Charnley 2003]{rodgers03a}; \cite[Lee et al. 2004]{lee04a}; \cite[Aikawa et al. 2008]{aikawa08a}), but such models cannot properly describe the circumstellar disk. We have recently constructed a new collapse model that describes the envelope and the disk in two dimensions (2D) as an axisymmetric structure (\cite[Visser et al. 2009a]{visser09a}, \cite[2011]{visser11a}; \cite[Visser \& Dullemond 2010]{visser10a}). By taking the vertical extent of the disk into account, we find that accretion from the envelope onto the disk largely takes place near the cold outer edge, rather than near the warmer inner parts as suggested by earlier models. The lower temperatures in our scenario allow for a larger fraction of pristine material to enter the disk. The 2D models also show strong evidence of layered accretion in the disk: new material arriving from the envelope always accretes on top of the older material already in the disk (Fig.\ \ref{fig:acclayers}). Another improvement in our model is that we calculate the dust temperatures throughout the disk and the envelope with a full radiative transfer code. The approximate temperature calculations used by some older models are not accurate enough for a reliable investigation of the chemistry.

We provide here a simple summary of our model (Sect.\ \ref{sec:mod}) and highlight some of the first results (Sect.\ \ref{sec:res}). In particular, we focus on the chemical history of H$_2$O and other simple molecules as material flows from the pre-stellar core to the planet-forming disk. The discussion (Sect.\ \ref{sec:disc}) addresses two open questions in star formation: how strong is the disk-envelope accretion shock, and do hot corinos really exist?


\section{Model description}
\label{sec:mod}
In order to follow the chemical evolution from a pre-stellar core to a circumstellar disk, we need a physical framework to provide densities, temperatures and other quantities that may affect the chemistry. We choose to use a fast semi-analytical model for the density and velocity profiles and a full continuum radiative transfer code for the dust temperature and the radiation field. The dust temperature is particularly important for the gas-grain interactions. Hydrodynamical codes often use an approximation of the dust temperature and are therefore less suited to study the chemical evolution.

\begin{figure}[t]
\begin{center}
\resizebox{\hsize}{!}{\includegraphics{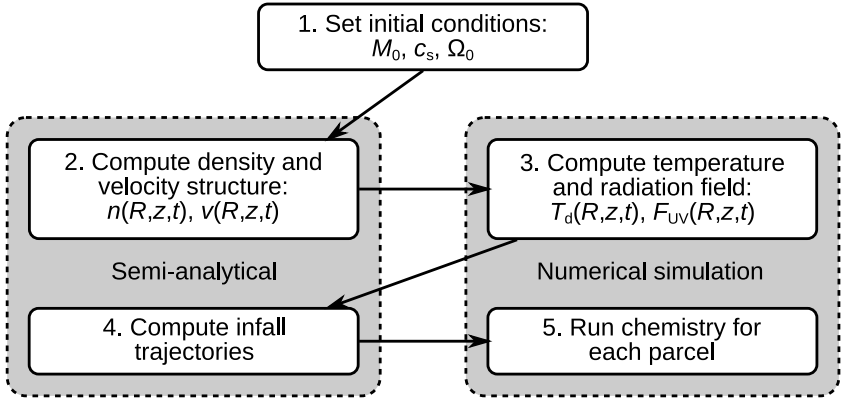}}
\caption{Step-wise summary of our 2D axisymmetric collapse model. Steps 2 and 4 are semi-analytical, while steps 3 and 5 consist of detailed numerical simulations. (Figure adapted from \cite[Visser et al. (2011)]{visser11a}.)}
\label{fig:msteps}
\end{center}
\end{figure}

The full model procedure is explained in detail in \cite[Visser et al. (2009a)]{visser09a}, \cite[Visser \& Dullemond (2010)]{visser10a} and \cite[Visser et al. (2011)]{visser11a}. We summarise the model in this section by way of the step-wise description from Fig.\ \ref{fig:msteps}.


\subsection{Initial conditions}
The physical and chemical evolution of the collapsing envelope depends strongly on the initial conditions. The three most prominent parameters in our model are the mass of the cloud core before the onset of collapse ($M_0$), the isothermal sound speed ($c_{\rm s}$), and the core's solid-body rotation rate ($\Omega_0$). A higher rotation rate leads to a larger and more massive disk, as do a lower sound speed and a higher core mass (\cite[Visser et al. 2009a]{visser09a}). Lowering the sound speed or increasing the core mass both tend to produce a colder disk, with larger fractions of ices relative to gas-phase species. Increasing the solid-body rotation rate also leads to a colder disk. The core mass and the sound speed together determine how long it takes for the entire core to accrete onto the star and disk: $t_{\rm acc} \equiv M_0/\dot{M} \approx M_0G/c_{\rm s}^3$ (\cite[Shu 1977]{shu77a}).


\subsection{Density and velocity}
We follow the inside-out solution of \cite[Shu (1977)]{shu77a} for a gravitationally collapsing spherical envelope. However, because of the core's rotation, spherical symmetry is lost at the onset of collapse. The effects of rotation can be treated as a perturbation of the pure spherical solution (\cite[Cassen \& Moosman 1981]{cassen81a}; \cite[Terebey et al. 1984]{terebey84a}). In the perturbed solution, matter no longer streams in radially, but it is deflected towards the gravitational midplane and builds up a circumstellar disk. After the disk is first formed, it evolves by ongoing accretion from the envelope and by viscous spreading to conserve angular momentum (\cite[Shakura \& Sunyaev 1973]{shakura73a}; \cite[Lynden-Bell \& Pringle 1974]{lyndenbell74a}; \cite[Dullemond et al. 2006]{dullemond06a}).

The size of the disk is initially set by the centrifugal radius ($R_{\rm c}\approx(1/16)c_{\rm s}^3t^3\Omega_0$, with $t$ the time since the onset of collapse), because the streamlines from the envelope only hit the midplane inside of that point. However, viscous spreading can push the outer edge of the disk to several times $R_{\rm c}$ (\cite[Visser et al. 2009a]{visser09a}; \cite[van Weeren et al. 2009]{vanweeren09a}). We adopt a Gaussian profile for the vertical density structure of the disk, from which we also derive the vertical velocity profile. The boundary between the disk and the envelope is conveniently defined as the surface where the thermal pressure from the disk equals the ram pressure from the infalling envelope (\cite[Visser \& Dullemond 2010]{visser10a}).

The analytical collapse solution of \cite[Terebey et al. (1984)]{terebey84a} does not include a bipolar outflow, so we have to put one in as an ad-hoc addition. Based on observations and theoretical predictions (\cite[Velusamy \& Langer 1998]{velusamy98a}; \cite[Cant{\'o} et al. 2008]{canto08a}), we adopt curved cavity walls whose outflow angle grows with time; specifically, at a height $z$ above the midplane, the size $R_{\rm cav}$ of the cavity is proportional to $t^2z^{2/3}$ (\cite[Visser et al. 2011]{visser11a}).


\subsection{Temperature and radiation field}
The chemical evolution depends strongly on the temperature, so we employ a full radiative transfer method instead of an analytical approximation for step 3 of our model. The only heating source taken into account is the central star, whose luminosity and effective temperature evolve according to \cite[Adams \& Shu (1986)]{adams86a} and \cite[Young \& Evans (2005)]{young05a}. Taking the density profiles from step 2, we compute the 2D dust temperature profiles at a range of time steps with RADMC (\cite[Dullemond \& Dominik 2004]{dullemond04a}). As additional output, RADMC also provides the intensity of the radiation field as function of wavelength at each point in the disk and remnant envelope, as required for the photon-driven reactions in our chemical network.

We set the gas temperature equal to the dust temperature throughout the disk and the envelope, even though this is a poor assumption in the surface of the disk and the inner parts of the envelope (\cite[Kamp \& Dullemond 2004]{kamp04a}; \cite[Jonkheid et al. 2004]{jonkheid04a}). Planned updates to the model include a separate treatment of the dust and gas temperatures. The effect of the accretion shock on the temperature profiles is discussed in Sect.\ \ref{sec:disc}.

\subsection{Infall trajectories}
We follow the chemical evolution as material streams from the natal core towards the circumstellar disk, so we have to solve the chemistry in a Lagrangian frame. The semi-analytical nature of the density and velocity solutions offers two ways to approach the problem. The first way, followed in \cite[Visser et al. (2009a)]{visser09a}, consists of populating the core with a large number of points (at least 10\,000) and calculating the infall streamlines in the forward direction. Alternatively, as done in \cite[Visser et al. (2011)]{visser11a}, we can select points of interest in the disk at the end of the collapse phase and calculate the trajectories backwards in time to their point of origin in the core. In both cases, we end up with a set of trajectories (Cartesian or polar coordinates as function of time) with corresponding physical conditions (density, temperature, UV field and extinction as function of position and time). These are used as input for the chemistry code. Each trajectory is treated in isolation, i.e., the material streaming along one trajectory is assumed not to interact with material streaming along other trajectories.


\subsection{Chemistry}
The basis of our chemical network is the UMIST06 database (\cite[Woodall et al. 2007]{woodall07a}) as modified by \cite[Bruderer et al. (2009)]{bruderer09a}, except that X-ray chemistry is not included. Full details are provided in \cite[Visser et al. (2011)]{visser11a}. The network includes freeze-out and evaporation (both thermal and photon-driven), but we limit the grain-surface chemistry to a handful of simple hydrogenation reactions: atomic H, C, N, O and S can react with incoming H to form H$_2$, CH$_4$, NH$_3$, H$_2$O and H$_2$S (\cite[Black \& van Dishoeck 1987]{black87a}; \cite[Bergin \& Langer 1997]{bergin97b}; \cite[Hollenbach et al. 2009]{hollenbach09a}). Other potentially important grain-surface reactions, such as the formation of H$_2$O out of O$_2$ (\cite[Ioppolo et al. 2008]{ioppolo08a}) or the formation of H$_2$CO and CH$_3$OH out of CO (\cite[Fuchs et al. 2009]{fuchs09a}), have not yet been incorporated. The gas-phase network includes photodissociation and photoionisation reactions (accounting specifically for self-shielding of CO; \cite[Visser et al. 2009b]{visser09b}), which dominate the chemistry close to the outflow wall and in the surface layers of the disk.

In order to set the chemical composition at the onset of collapse ($t=0$), we evolve the initially atomic gas for a period of 1 Myr at $n_{\rm H}=8\times10^{4}$ cm$^{-3}$ and $T_{\rm gas}=T_{\rm dust}=10$ K\@. The resulting solid and gas-phase abundances are consistent with those observed in pre-stellar cores (e.g., \cite[di Francesco et al. 2007]{difrancesco07a}). They form the initial conditions for the chemical evolution along each infall trajectory.


\section{Results}
\label{sec:res}


\subsection{Key chemical trends}
This section highlights a few key results from the standard model of \cite[Visser et al. (2009a]{visser09a}, \cite[2011)]{visser11a}, which has initial conditions $M_0=1.0$ $M_\odot$, $c_{\rm s}=0.26$ km s$^{-1}$ and $\Omega_0=10^{-14}$ s$^{-1}$. The chemistry is run for the full duration of the envelope accretion phase ($t_{\rm acc}=2.52\times10^{5}$ yr), following the 1 Myr pre-collapse phase.

\begin{figure}[t]
\begin{center}
\resizebox{\hsize}{!}{\includegraphics{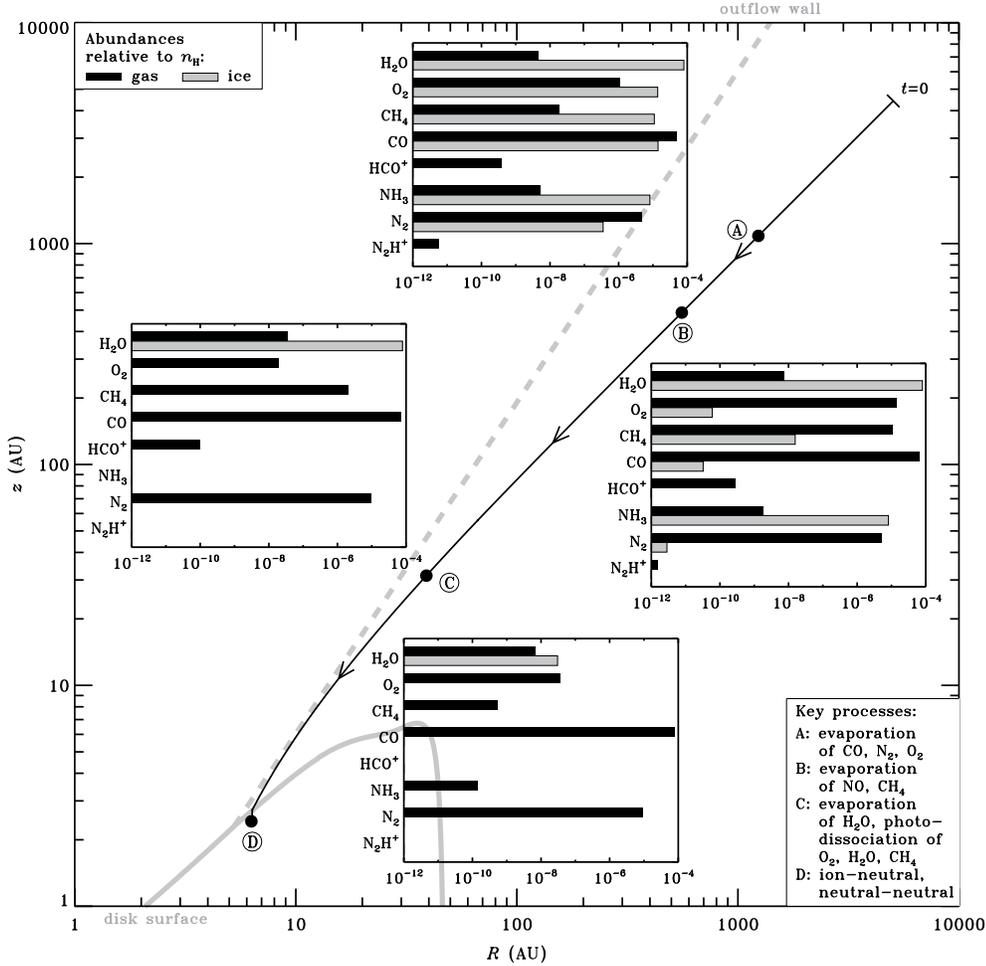}}
\caption{Schematic view of the chemical evolution along one infall trajectory (solid black curve). The solid and dashed grey lines denote the surface of the disk and the outflow wall, both at $t=t_{\rm acc}$. The abundances of some key species (black bars: gas; grey bars: ice) are plotted at four points along the trajectory (A: top panel; B: right panel; C: left panel; D: bottom panel). The key processes governing the overall chemistry at each point are listed in the bottom right. (Figure adapted from \cite[Visser et al. (2011)]{visser11a}.)}
\label{fig:trajchem}
\end{center}
\end{figure}

We first focus on the chemical evolution along one particular trajectory, starting near the edge of the original core ($R_{\rm i}=5050$ AU, $z_{\rm i}=4420$ AU) and ending at $t=t_{\rm acc}$ in the inner part of the disk, about 0.2 AU below the surface ($R_{\rm f}=6.3$ AU, $z_{\rm f}=2.4$ AU). The chemistry along this and other trajectories is governed by changes in the physical conditions. Figure \ref{fig:trajchem} shows how the solid and gas-phase abundances of eight important species change along the streamline. We identify four chemical events:
\begin{itemize}
\item[$\bullet$] At point A, the temperature reaches a value of about 20 K and simple ices like CO, N$_2$ and O$_2$ begin to evaporate.
\item[$\bullet$] The temperature increases further to 30--40 K at point B, driving CH$_4$ and NO off the grains.
\item[$\bullet$] Moving in further, the dust is heated to the evaporation temperature of H$_2$O (100 K). This close to the outflow cavity wall, the gas is exposed to the stellar UV field, dissociating the evaporating H$_2$O and other previously evaporated species. Our model does not include an excess UV field for the young star, so there are not enough high-energy photons to dissociate CO.
\item[$\bullet$] Finally, at point D, the high density ($10^{10}$--$10^{11}$ cm$^{-3}$) allows some of the dissociated species to reform in ion-neutral and neutral-neutral reactions.
\end{itemize}


\subsection{Chemical stratification}
Different streamlines exhibit different physical conditions. In order to fully understand the chemical evolution from every point in the pre-stellar core to every point in the circumstellar disk, we have to investigate more than just the single trajectory from Fig.\ \ref{fig:trajchem}. In this regard, it is interesting to look at where material from different parts of the core ends up in the disk.

The hydrodynamical simulations of \cite[Brinch et al. (2008)]{brinch08a} and our semi-analytical model (\cite[Visser et al. 2009a]{visser09a}) both show the envelope to accrete onto the disk in a layered fashion (Fig.\ \ref{fig:acclayers}). Material from near the centre of the pre-stellar core is the first to become part of the disk, and material from larger radii continuously piles up on top. The thickness of the disk is small relative to its radial size (typical scale heights are 0.1--0.2$R$), but the outer edge is high enough that a substantial fraction of the infalling material hits there rather than on the surface (Fig.\ \ref{fig:acclayers}; see also Fig.\ 3 in \cite[Visser \& Dullemond 2010]{visser10a}). The infalling material has to compete with the outwardly expanding disk, resulting in some material being pushed up and over the outer edge in the hydrodynamical simulations (see also \cite[van Weeren et al. 2009]{vanweeren09a}).

\begin{figure}[t]
\begin{center}
\resizebox{0.75\hsize}{!}{\includegraphics{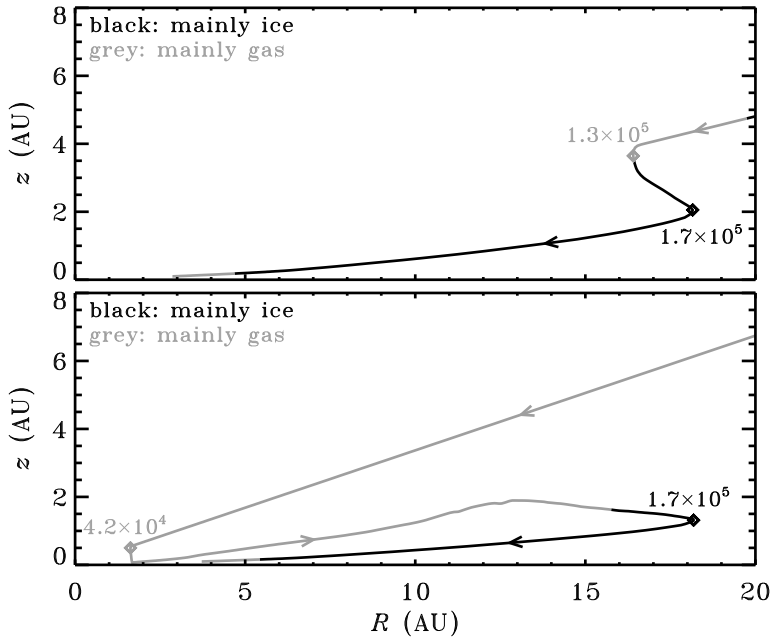}}
\caption{Examples of qualitatively different infall trajectories terminating at nearly the same point at the end of the accretion phase: $R=2.9$ AU (top) and 3.7 AU (bottom). The black and grey portions of the trajectories indicate where most of the water is frozen out or in the gas phase. The points where the trajectories reverse direction are marked with the time in years.}
\label{fig:traj}
\end{center}
\end{figure}

Individual streamlines can exhibit large qualitative differences despite terminating at nearly the same point in the disk. Figure \ref{fig:traj} shows two such streamlines, with termini at 2.9 and 3.7 AU from the star. The streamline in the bottom panel originates at $r=1100$ AU from the centre of the pre-stellar core. It reaches the disk at a fairly early time ($4.2\times10^{4}$ yr after the onset of collapse), when the disk is only a few AU large. The parcel of gas following this streamline gets swept up in the outward viscous expansion until, at $t=1.7\times10^{5}$ yr, enough mass has accreted at larger radii that it gets pushed back in again. The streamline in the bottom panel originates further out in the natal core, at $r=3500$ AU, and therefore accretes at a later time, $t=1.3\times10^{5}$ yr, when the disk has grown to about 20 AU\@. It experiences a brief period of outward motion before moving in further.

In terms of the chemical evolution of the material flowing along these two streamlines, the key difference lies in the conditions encountered before $t=10^5$ yr. The bottom streamline passes much closer to the star than does the other one (1.6 versus 16 AU); it is heated to more than 500 K, whereas the top parcel initially does not get warmer than 120 K\@. The latter is enough to drive H$_2$O and other non-refractory species off the grains (as indicated by the grey portion of the streamline up to $1.3\times10^{5}$ yr), but not enough to power the kind of high-temperature gas-phase chemistry that is possible at 500 K\@.

\begin{figure}[t]
\begin{center}
\resizebox{0.8\hsize}{!}{\includegraphics{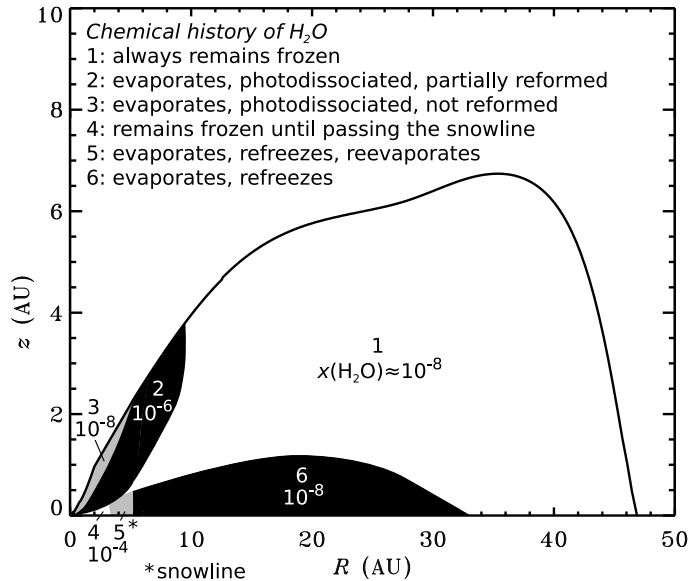}}
\caption{Schematic view of the history of H$_2$O gas and ice towards different parts of the disk. The typical gas-phase abundance is indicated for each zone (zones 4 and 5 both have $10^{-4}$). The position of the snowline ($R=5$ AU, between zones 5 and 6) is marked with an asterisk. Note the disproportionality of the $R$ and $z$ axes. (Figure adapted from \cite[Visser et al. (2011)]{visser11a}.)}
\label{fig:wzones}
\end{center}
\end{figure}

The different types of streamlines give rise to different chemical zones in the disk. This effect of ``chemical stratification'' is visualised in Fig.\ \ref{fig:wzones}, which traces the history of H$_2$O gas and ice towards various parts of the disk. (Similar figures for CH$_4$ and NH$_3$ appear in \cite[Visser et al. (2011)]{visser11a}.) The streamlines terminating in zone 1 never encounter temperatures above 100 K, so H$_2$O always remains frozen on the grains. Zone 2 contains the terminus of the streamline from Fig.\ \ref{fig:trajchem}; some H$_2$O ends up here, but most of it is dissociated by the stellar radiation field. The same happens for material ending up in zone 3, except that no H$_2$O reforms at all after being destroyed. Parcels ending up in zone 4 do not encounter a strong radiation field; H$_2$O in these parcels stays on the grains until passing the snowline at $\sim$5 AU. Finally, zones 5 and 6 consist largely of material that accreted at early times, following trajectories like the one in the bottom panel of Fig.\ \ref{fig:traj}. H$_2$O in these parcels undergoes a freeze-thaw cycle: it evaporates during the initial inwards part of the trajectory and freezes out again during the outwards leg. Parcels ending up in zone 5 undergo another evaporation event as they pass the snowline; material in zone 6 lies beyond the snowline, so H$_2$O remains frozen the rest of the time.

The chemical stratification comes with one major caveat: mixing. In our model, there is no exchange of material between adjacent parcels. In reality, however, turbulent mixing is likely to change the sizes and locations of the chemical zones, and to make the borders between the zones more diffuse (e.g., \cite[Semenov et al. 2006]{semenov06a}). Spectroscopic observations at 5--10 AU resolution, as possible with ALMA, are required to determine to what extent chemical stratification occurs in actual disks.


\section{Discussion}
\label{sec:disc}


\subsection{How strong is the disk-envelope accretion shock?}
\label{sec:accshock}
When the supersonically infalling material from the envelope hits the nearly static disk, it passes through a J-type shock. However, this shock may not play as large a role as commonly believed. \cite[Neufeld \& Hollenbach (1994)]{neufeld94a} showed that the strength of the shock decreases with decreasing pre-shock density and decreasing shock velocity. For a given radius, the density and velocity decrease as the collapse proceeds; for a given point in time, both decrease with radius (\cite[Cassen \& Moosman 1981]{cassen81a}; \cite[Terebey et al. 1984]{terebey84a}).

During the second half of the collapse phase (from $t=0.5t_{\rm acc}$ to $t_{\rm acc}$), at least 80\% of the infalling material accretes onto the outer half of the disk. (The exact number depends on the initial conditions.) At the end of the collapse phase, the bulk of the disk consists of material that accreted at fairly late stages and large radii.  Most of the material that accreted at early times or small radii gets pushed into the star. Hence, we expect that the bulk of the gas and the dust in a T Tauri or Herbig Ae/Be disk did not pass through a strong accretion shock. Quantitatively, we showed in \cite[Visser et al. (2009a)]{visser09a} that the shock has a smaller effect on the temperature history of the disk material than does the radiation from the young star. In terms of the chemical evolution, this means the desorption of H$_2$O and organic species is likely unaffected by the accretion shock. Species with lower binding energies, however, may be more sensitive.


\subsection{Do hot corinos exist?}
Observations of complex organic species like methyl formate (HCOOCH$_3$) and dimethyl ether (CH$_3$OCH$_3$) in low-mass protostars have been linked to the high-temperature ($>100$ K) inner envelope, also known as the hot core or hot corino (\cite[Ceccarelli 2004]{ceccarelli04a}). One possible formation pathway is powered by the evaporation of H$_2$O and other non-refractory species at 100 K, followed by a chain of high-temperature gas-phase reactions to transform simple carbon-bearing molecules into second-generation complex organics (\cite[Herbst \& van Dishoeck 2009]{herbst09a}; see also \cite[Charnley et al. 1992]{charnley92a}). Alternatively, the organics may be formed abundantly on the grain surfaces through mild photochemistry at 20--40 K (\cite[Garrod \& Herbst 2006]{garrod06a}; \cite[Garrod et al. 2008]{garrod08a}). Which scenario is more likely?

The first formation route requires the gas to spend at least several $10^{3}$ yr at temperatures above 100 K (\cite[Charnley et al. 1992]{charnley92a}). In our model, however, the material in the inner envelope is essentially in freefall towards the star or the inner disk, and it only spends a few 100 yr at the required temperatures (\cite[Visser et al. 2009a]{visser09a}; see also \cite[Sch{\"o}ier et al. 2002]{schoier02a}). The grain-surface pathway takes several $10^{4}$ yr at 20--40 K (\cite[Garrod \& Herbst 2006]{garrod06a}). Such temperatures are attainable both in the envelope before the material reaches the hot inner part and inside the disk (e.g., Fig.\ 12 in \cite[Visser et al. 2009a]{visser09a}). Once formed on the grains, these first-generation organics can be returned to the gas phase by thermal evaporation in the hot corino, the hot inner disk, or the surface layers of the outer disk. They can also be liberated by non-thermal processes such as photodesorption in the disk and shocks in the outflow walls. In any scenario, the mass of the hot corino is quite limited: the amount of material wedged between the disk and the outflow cavity is at most a few per cent of the amount of material inside the disk. Hence, even if the hot corino is rich in organics, there still would not be enough of them to explain the observations.

Based on the timescales and the mass, we conclude that the observed second-generation complex organics are not associated with the hot core or hot corino. The circumstellar disk and outflow walls offer a more likely explanation: they have a larger mass reservoir and are dynamically stable enough for the observed organics to form abundantly.


\section{Conclusions}
We have constructed a two-dimensional axisymmetric model to simulate the collapse of a pre-stellar core into a young star and its surrounding disk. We use the model to study the chemical evolution of the gas and the dust streaming from the envelope to the disk. Comparing the streamlines to what happens in one-dimensional models, we conclude that it is imperative to study core collapse and disk formation in 2D\@. Our model predicts chemical stratification in the disk due to different physical conditions (density, temperature, radiation field) encountered along different streamlines. We argue that the disk-envelope accretion shock does not play a significant role for the material in the disk at the end of the collapse phase. Finally, our model suggests that complex organic species are formed on the grain surfaces at temperatures of 20 to 40 K, rather than in the gas phase in the $T>100$ K hot corino.

\begin{discussion}

\discuss{Yamamoto}{How do you assume the initial conditions in your model? We now know that various complex molecules exist in protostellar cores. How are they brought into the inner disk?}

\discuss{Visser}{We have a static pre-collapse phase of $10^6$ yr to produce typical pre-stellar core abundances. Since we do not have any grain-surface chemistry other than simple hydrogenation of O, C, N and S, complex organics are not formed abundantly. However, based on the models by \cite[Garrod \& Herbst (2006)]{garrod06a} and \cite[Garrod et al. (2008)]{garrod08a}, we predict that complex organics are formed abundantly inside the disk. We have not considered warm carbon-chain chemistry.}

\discuss{Johnstone}{How important are accretion shocks for the chemical evolution of infalling gas?}

\discuss{Visser}{The accretion shock is not important for the bulk of the material in the disk at the end of the accretion phase, because that material accreted at late times and large radii, and therefore did not encounter a strong shock (\cite[Neufeld \& Hollenbach 1994]{neufeld94a}). Material that accretes earlier encounters a stronger shock, but this material disappears into the star while the rest of the disk is formed. We refer to Sect.\ \ref{sec:accshock} for more details.}

\discuss{Aikawa}{What is the cause of the vortices that appear in the animation? Do they also appear in your 2D analytical model?}

\discuss{Visser}{The vortices are due to material piling up at the outer edge of the disk and being pushed up and over the outer edge in the 2D simulation. They do not appear in the semi-analytical model.}

\discuss{Irvine}{Evidence suggests that most stars, including the Sun, did not form in isolation (the Shu model). Have you considered how this might affect your results?}

\discuss{Visser}{No, we have not considered this in any detail. We expect that many of the results hold for circumbinary envelopes and disks, as long as the interbinary distance is smaller than 10 AU or so. If the protostar in question forms in the vicinity of an O or B star, the strong external radiation field would result in a larger fraction of atomic and ionised species at the outer edge of the envelope and the disk.}

\end{discussion}


\begin{thebibliography}{}

\bibitem[]{adams86a}
{Adams, F.C. \& Shu, F.H.} 1986, \apj, 308, 836

\bibitem[]{aikawa08a}
{Aikawa, Y., Wakelam, V., Garrod, R.T., \& Herbst, E.} 2008, \apj, 674, 984

\bibitem[]{bergin97b}
{Bergin, E.A. \& Langer, W.D.} 1997, \apj, 486, 316

\bibitem[]{black87a}
{Black, J.H. \& van Dishoeck, E.F.} 1987, \apj, 322, 412

\bibitem[]{brinch08a}
{Brinch, C., Hogerheijde, M.R., \& Richling, S.} 2008, \aap, 489, 607

\bibitem[]{bruderer09a}
{Bruderer, S., Doty, S.D., \& Benz, A.O.} 2009, \apjs, 183, 179

\bibitem[]{canto08a}
{Cant{\'o}, J., Raga, A.C., \& Williams, D.A.} 2008, Rev.\ Mex.\ Astron.\ Astr., 44, 293

\bibitem[]{cassen81a}
{Cassen, P. \& Moosman, A.} 1981, \textit{Icarus}, 48, 353

\bibitem[]{ceccarelli04a}
{Ceccarelli, C.} 2004, in \textit{ASP Conf.\ Ser. 323: Star Formation in the Interstellar Medium: In Honor of David Hollenbach}, ed. D.\ Johnstone, F.C.\ Adams, D.N.C.\ Lin, D.A.\ Neufeld, \& E.C.\ Ostriker (San Francisco: ASP), 195

\bibitem[]{difrancesco07a}
{di Francesco, J., Evans, II, N.J., Caselli, et al.} 2007, \textit{Protostars \& Planets V}, 17

\bibitem[]{dullemond04a}
{Dullemond, C.P. \& Dominik, C.} 2004, \aap, 417, 159

\bibitem[]{dullemond06a}
{Dullemond, C.P., Apai, D., \& Walch, S.} 2006, \apjl, 640, L67

\bibitem[]{fuchs09a}
{Fuchs, G.W., Cuppen, H.M., Ioppolo, S., et al.} 2009, \aap, 505, 629

\bibitem[]{garrod06a}
{Garrod, R.T. \& Herbst, E.} 2006, \aap, 457, 927

\bibitem[]{garrod08a}
{Garrod, R.T., Weaver, S.L.W., \& Herbst, E.} 2008, \apj, 682, 283

\bibitem[]{herbst09a}
{Herbst, E. \& van Dishoeck, E.F.} 2009, \araa, 47, 427

\bibitem[]{hollenbach09a}
{Hollenbach, D., Kaufman, M.J., Bergin, E.A., \& Melnick, G.J.} 2009, \apj, 690, 1497

\bibitem[]{ioppolo08a}
{Ioppolo, S., Cuppen, H.M., Romanzin, C., van Dishoeck, E.F., \& Linnartz, H.} 2008, \apj, 686, 1474

\bibitem[]{jonkheid04a}
{Jonkheid, B., Faas, F.G.A., van Zadelhoff, G.-J., \& van Dishoeck, E.F.} 2004, \aap, 428, 511

\bibitem[]{kamp04a}
{Kamp, I. \& Dullemond, C.P.} 2004, \apj, 615, 991

\bibitem[]{lee04a}
{Lee, J.-E., Bergin, E.A., \& Evans, II, N.J.} 2004, \apj, 617, 360

\bibitem[]{lyndenbell74a}
{Lynden-Bell, D. \& Pringle, J.E.} 1974, \mnras, 168, 603

\bibitem[]{neufeld94a}
{Neufeld, D.A. \& Hollenbach, D.J.} 1994, \apj, 428, 170

\bibitem[]{rodgers03a}
{Rodgers, S.D. \& Charnley, S.B.} 2003, \apj, 585, 355

\bibitem[]{schoier02a}
{Sch{\"o}ier, F.L., J{\o}rgensen, J.K., van Dishoeck, E.F., \& Blake, G.A.} 2002, \aap, 390, 1001

\bibitem[]{semenov06a}
{Semenov, D., Wiebe, D., \& Henning, T.} 2006, \apjl, 647, L57

\bibitem[]{shu77a}
{Shu, F.H.} 1977, \apj, 214, 488

\bibitem[]{terebey84a}
{Terebey, S., Shu, F.H., \& Cassen, P.} 1984, \apj, 286, 529

\bibitem[]{vanweeren09a}
{van Weeren, R.J., Brinch, C., \& Hogerheijde, M.R.} 2009, \aap, 497, 773

\bibitem[]{velusamy98a}
{Velusamy, T. \& Langer, W.D.} 1998, \nat, 392, 685

\bibitem[]{visser09a}
{Visser, R., van Dishoeck, E.F., Doty, S.D., \& Dullemond, C.P.} 2009a, \aap, 495, 881

\bibitem[]{visser09b}
{Visser, R., van Dishoeck, E.F., \& Black, J.H.} 2009b, \aap, 503, 323

\bibitem[]{visser10a}
{Visser, R. \& Dullemond, C.P.} 2010, \aap, 519, A28

\bibitem[]{visser11a}
{Visser, R., Doty, S.D., \& van Dishoeck, E.F.} 2011, \aap, submitted

\bibitem[]{woodall07a}
{Woodall, J., Ag{\'u}ndez, M., Markwick-Kemper, A.J., \& Millar, T.J.} 2007, \aap, 466, 1197

\bibitem[]{young05a}
{Young, C.H. \& Evans, II, N.J.} 2005, \apj, 627, 293

\end{thebibliography}
\end{document}